\documentclass{elsart41}
\usepackage{graphics}
\usepackage{graphicx}
\usepackage{amssymb}
\begin{document}

\begin{frontmatter}



\title{Spatially resolved NMR relaxation rate
in a noncentrosymmetric superconductor}
%---- Don't remove this comment line! ----

%\author[AA]{N. Hayashi\corauthref{Name1}},
%\ead{hayashi@itp.phys.ethz.ch}
\author[AA]{N. Hayashi},\,
\author[AA]{K. Wakabayashi},\,
\author[AA]{P. A. Frigeri},\,
\author[CC]{Y. Kato},\,
\author[AA]{M. Sigrist}

\address[AA]{Institut f\"ur Theoretische Physik,
ETH-H\"onggerberg,
CH-8093 Z\"urich, Switzerland}  
%\address[BB]{Department of Quantum Matter Science,
%Graduate School of Advanced Sciences of Matter (ADSM),
%Hiroshima University, Higashi-Hiroshima 739-8530, Japan}
\address[CC]{Department of Basic Science,
University of Tokyo,
Tokyo 153-8902, Japan}

%\corauth[Name1]{Corresponding author. Fax: +41-1-633-1115}

\begin{abstract}
   We numerically study the spatially-resolved
%nuclear magnetic resonance
NMR
around a single vortex
in a noncentrosymmetric superconductor such as CePt$_3$Si.
   The nuclear spin-lattice relaxation rate $T_1^{-1}$
under the influence of the vortex core states is calculated
for an $s+p$-wave Cooper pairing state.
   The result is compared with that for an $s$-wave pairing state.
\end{abstract}

\end{frontmatter}

% main text

   The heavy fermion superconductor CePt$_{3}$Si has
a noncentrosymmetric crystal structure \cite{bauer}.
   The lack of the inversion symmetry
leads to the mixture of Cooper pairing channels of different parity,
resulting in unusual properties observed
experimentally \cite{bauer,bauer3,bauer2}.
   For example, it is observed that
while a Hebel-Slichter coherence peak
appears
in the nuclear spin-lattice relaxation rate
$T_{1}^{-1}$ \cite{bauer3,bauer2,yogi}
suggesting an $s$-wave like Cooper pairing,
low-temperature experiments on
the London penetration depth \cite{bauer2,bonalde}
and the thermal conductivity \cite{izawa}
indicate line nodes in the quasiparticle gap
(i.e., unconventional superconductivity).
   In this context,
$T_{1}^{-1}$ was recently discussed
in Refs.\ \cite{fujimoto,hayashi,samokhin} theoretically.

   On the other hand,
the spatially-resolved nuclear magnetic resonance (NMR) method
has been revealed to be
a powerful experimental technique \cite{takigawa,kakuyanagi}.
   This technique as a probe of the electronic structure
with spatial resolution
is expected to reveal pairing symmetry
of unconventional superconductors,
because in spatially inhomogeneous systems there appear
properties specific to the unconventional superconductivity.

   In this paper, we study
the spatially-resolved $T_1^{-1}$
around a single vortex
for an $s+p$-wave pairing state.
   This pairing state is proposed for CePt$_{3}$Si
in Ref.\ \cite{paolo3}.
   We consider a system described in Refs.\ \cite{paolo3,paolo1}.
   From now on, the notations are the same as
those in Ref.\ \cite{hayashi}.

   To obtain $T_1^{-1}$ in the spatially inhomogeneous situation,
we calculate the quasiclassical Green function ${\check g}$
which is a $4 \times 4$ matrix in Nambu and spin spaces.
   ${\check g}$ follows the Eilenberger equation
for the noncentrosymmetric superconductivity \cite{hayashi,hayashi2}
%================================
\begin{equation}
\vspace{-1mm}
i {\mathbf v}_{\mathrm F} \cdot
{\mathbf \nabla}{\check g}
+ \bigl[ i\omega_n {\check \tau}_{3}
-\alpha {\check {\mathbf g}}_k \cdot {\check {\mathbf S}}
-{\check \Delta}_k,
{\check g} \bigr]
=0,
\label{eq:eilen0}
\vspace{-1mm}
\end{equation}
%================================
where $\alpha$ indicates the strength of the Rashba-type spin-orbit coupling,
${\check {\mathbf g}}_k$ is composed of
the antisymmetric vector ${\mathbf g}_k=(-{\tilde k}_y,{\tilde k}_x,0)$
\cite{paolo1},
${\check {\mathbf S}}$ is the electron spin operator,
and ${\check \Delta}_k$ is the superconducting order parameter.
   For a static magnetic field in the $z$ direction,
the expression for $1/T_1 T$ is given as \cite{hayashi,hayashi3}
%================================
\begin{eqnarray}
\vspace{-1mm}
\frac{T_1(T_{\mathrm c}) T_{\mathrm c}}{T_1(T) T}
=
\frac{1}{4T}
\int^{\infty}_{-\infty}
\frac{d \omega}{\cosh^2(\omega/2T)}
W(\omega),
\vspace{-1mm}
\end{eqnarray}
%================================
%================================
\begin{eqnarray}
\vspace{-1mm}
W(\omega)
=
\langle a^{22}_{\downarrow \downarrow}(\omega) \rangle
\langle a^{11}_{\uparrow \uparrow}(-\omega) \rangle
-
\langle a^{21}_{\downarrow \uparrow}(\omega) \rangle
\langle a^{12}_{\uparrow \downarrow}(-\omega) \rangle,
%\nonumber \\
\vspace{-1mm}
\end{eqnarray}
%================================
%================================
\begin{eqnarray}
\vspace{-1mm}
{\check a}(\omega)
%&=&       \nonumber
=
\frac{i}{2\pi} {\check \tau}_3
\bigl[
   {\check g}(
      i\omega_n \rightarrow \omega +i\eta)
%   \\
%   & & {}
%   \qquad \quad
   -
   {\check g}(
      i\omega_n \rightarrow \omega -i\eta)
\bigr],
\end{eqnarray}
%================================
where
${\check a} = ({\hat a}^{ij})$
and
${\hat a}^{ij} = (a^{ij}_{\mu\nu})$,
[$i,j=\{1,2\}$, $\mu,\nu=\{\uparrow, \downarrow\}$].
   The brackets $\langle \cdots \rangle$ denote
the average over the Fermi surface.
$\eta$ ($> 0$) is a small constant,
which can roughly represent the impurity effect.

   We consider two Fermi surfaces split by
the spin-orbit coupling $\alpha {\mathbf g}_k \cdot {\mathbf S}$,
and have the $s+p$-wave order parameter
$\Psi+\Delta\sin\theta$ on one Fermi surface
and $\Psi-\Delta\sin\theta$ on the other one \cite{paolo3}
($\Psi$: $s$-wave, $\Delta$: $p$-wave).
   We assume that
the difference of the density of states and $v_{\mathrm F}$
between the two Fermi surfaces
is small,
and neglect it here.
   We use the following empirical form of a single vortex,
%%================================
$\Phi/T_{\mathrm c}=a F(t) \tanh(r/\xi(t)) \exp(i\phi_r)$,
$\xi(t)=\xi_0 \sqrt{t} / b F(t)$,
$F(t)=\tanh(1.74 \sqrt{1/t-1})$.
%================================
Here,
$\Phi=\{ \Psi,\Delta \}$,
$\xi_0 = v_{\mathrm F}/T_{\mathrm c}$,
$t=T/T_{\mathrm c}$
and the fitting parameters $a$, $b$.
   The flux line is along the $z$ axis,
$r$ is the distance from the vortex center,
and $\phi_r$ is the angle in the $x$-$y$ plane.
   We have performed a full numerical computation as in Ref.\ \cite{hayashi4}
and obtained a self-consistent vortex solution for the same parameters
used in Ref.\ \cite{hayashi}.
   We then fitted it
with the above expression, and obtained
$(a,b)=(0.68,2.2)$ for $\Psi$ and $(1.42,2.2)$ for $\Delta$.
   We have also done the same fitting
for the conventional $s$-wave pairing case, so that $(a,b)=(1.76,2.2)$.
   We found that the fitting is good for $T > 0.1T_{\mathrm c}$,
while for $T \le 0.1T_{\mathrm c}$
it gives rather smaller core radius than actual one.

   We calculate $1/T_1 T$ at several positions around the vortex,
and show the results in Fig.\ 1 ($s+p$-wave state)
and Fig.\ 2 ($s$-wave state).
   Away from the vortex ($r=10\xi_0$),
$1/T_1 T$ exhibits the coherence peak just below $T_{\mathrm c}$
in both pairing states, while
it is rather smaller in the $s+p$-wave state \cite{hayashi}.
   With decreasing $r$ up to $\xi_0$,
the height of the peak gradually decreases.
   With further decreasing $r$ below $\xi_0$ inside the vortex core,
the height of the peak increases again and the value of $1/T_1 T$
at low temperatures increases also.
   The difference between $1/T_1 T$ at $r=0.6 \xi_0$ and that at $r=10\xi_0$
in Fig.\ 1 ($s+p$-wave state)
is similar to the difference between higher and lower frequency data
of experimentally observed $1/T_1 T$ in Refs.\ \cite{bauer3,bauer2,yogi}.
   That is, almost same hight of the peak just below $T_{\mathrm c}$
inside and outside the core, and larger values of $1/T_1 T$
at low temperatures
inside the core than those outside the core.
   Finally, we note that
$1/T_1 T$
increases further and the peak shifts to lower temperature side
with approaching the vortex center
(see plots for $r=0.2 \xi_0$ in Figs.\ 1 and 2).
   The difference between $s+p$-wave state (Fig.\ 1)
and $s$-wave state (Fig.\ 2) is conspicuous there
in the low-temperature behavior.
   It would be due to a difference of the vortex core states
between these pairing states.

   In conclusion, we calculated
spatially-resolved $1/T_1 T$
for $s+p$-wave and $s$-wave pairing states.
   The information on $1/T_1 T$ deeply inside the core might be
valuable for experimentally detecting
the pairing symmetry for a noncentrosymmetric superconductor.
%
%%%%%%%%%%%%%%%%%%%%%%%%%%%%%%%%%%%%%%%%%%%%%%%%%%%%%%%%%%%%%%%%%%%%%%%%%
\begin{figure}[!ht]
\begin{center}
\includegraphics[width=0.45\textwidth]{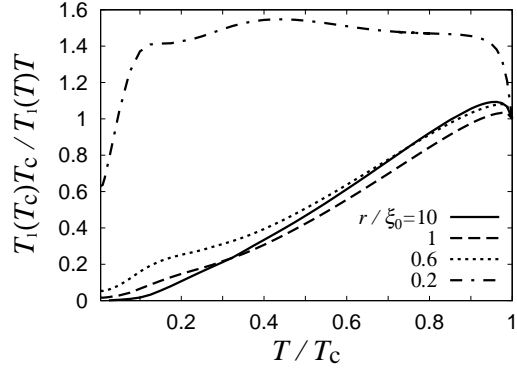}
\end{center}
\caption{
   $1/T_1 T$ vs $T$ for the $s+p$-wave pairing state.
   $r$ is the distance from the vortex center.
   $\eta = 0.01 T_{\mathrm c}$.
}
\label{fig1}
\end{figure}
%%%%%%%%%%%%%%%%%%%%%%%%%%%%%%%%%%%%%%%%%%%%%%%%%%%%%%%%%%%%%%%%%%%%%%%%%
%%%%%%%%%%%%%%%%%%%%%%%%%%%%%%%%%%%%%%%%%%%%%%%%%%%%%%%%%%%%%%%%%%%%%%%%%
\begin{figure}[!ht]
\begin{center}
\includegraphics[width=0.45\textwidth]{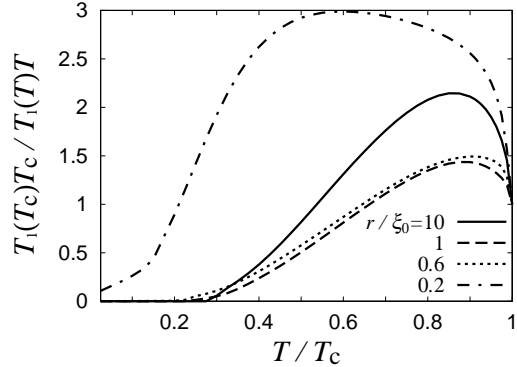}
\end{center}
\caption{
   $1/T_1 T$ vs $T$ for the $s$-wave pairing state.
   $\eta = 0.01 T_{\mathrm c}$.
}
\label{fig2}
\end{figure}
%%%%%%%%%%%%%%%%%%%%%%%%%%%%%%%%%%%%%%%%%%%%%%%%%%%%%%%%%%%%%%%%%%%%%%%%%
%

   We are grateful for financial support
from SNF and NCCR MaNEP.
   N.H.\ is also supported by
2003 JSPS Postdoctoral Fellowships for Research Abroad.

\end{document}